\documentclass[twocolumn,preprintnumbers]{revtex4-1}

\usepackage{graphicx}
\usepackage{epsf}
\usepackage{color}
\usepackage{amsmath,amssymb}
\usepackage{ascmac}
\usepackage{bm}
\usepackage{color}
\usepackage{ulem}

\usepackage{version}	% required for `\comment' (yatex added)
\begin{document}

\title{Lattice QCD study of static quark and antiquark correlations at finite $T$ via entanglement entropies}

\author{Toru T. Takahashi}
\affiliation{National Institute of Technology, Gunma college, Maebashi, Gunma
371-8530, Japan}
\author{Yoshiko Kanada-En'yo}
\affiliation{Department of Physics, Kyoto University, 
Sakyo, Kyoto 606-8502, Japan}

\date{\today}

 \begin{abstract}
  With the aim of clarification of color correlations among quarks,
  we investigate the color correlation between
  a static quark and an antiquark (static $q\bar q$) 
  below and above the phase transition temperature $T_c$
  through the entanglement entropy(EE).
  By a quenched lattice QCD calculation on an anisotropic lattice
  adopting the standard Wilson gauge action in the Coulomb gauge,
  we compute a reduced density matrix $\rho$ defined in the color space,
  and the entanglement entropy $S_{\rm EE}$ constructed from $\rho$.
  The spatial volume is $L^3 = 24^3$
  and the temporal extents are $N_T = 12,13,14,15,16,18,20$ and $24$,
  with a gauge coupling $\beta = 5.75$ and a renormalized 
  anisotropy $\xi = 4.0$,
  which corresponds to temperatures between 180 and 370 MeV.
  From an analysis of $\rho$ and $S_{\rm EE}$,
  the color correlation between $q\bar q$ pairs is obtained
  as a function of the interquark distance $R$ and the temperature $T$.
  Below $T_c$, the $R$-dependence of the color correlation
  resembles that at $T=0$:
  the quark's color correlation gradually decreases as $R$ increases
  due to the color screening by in-between gluons.
  Above $T_c$,
  due to the deconfinement phase transition,
  the color correlation among quarks is found to
  quickly disappear.
  Further,
  we investigate the color screening effect 
  via the screening mass defined by $\rho$, 
  and discuss the differences in the screening properties
  between the small and large R regions.
 \end{abstract}

\maketitle

\section{Introduction}
\label{Sec.Introduction}

Color confinement is one of the important nonperturbative features of Quantum
ChromoDynamics (QCD), which is the fundamental theory of the strong interaction.
In the confinement phase, no colored state can be isolated in vacuum;
hence, all the hadronic states are colorless.
This color confining feature of QCD has been studied and confirmed
using several approaches~\cite{Greensite:2011},
and it may be explained by a linear
potential among quarks arising in the large-separation limit (in the quenched QCD),
which has been extensively investigated by analyses of interquark potentials
for a variety of situations including multiquark systems~\cite{Bali:1994de,Takahashi:2002bw,Okiharu:2004ve}.
At the same time, above the phase transition temperature $T_c$, 
the confining potential among quarks vanishes, and colored states are allowed,
which has also been confirmed in lattice QCD calculations~\cite{Kaczmarek:2004gv,Burnier:2016mxc}.
Such confining potential may be illustrated by the flux-tube formation among quarks.
A color flux tube that has a constant energy per unit length
is formed among (totally color singlet) quarks, leading to the linearly rising
potential~\cite{Bali:1994de,Bornyakov:2004uv} in the large separation limit.

To cast light on the color structure of a confined quark pair,
in Ref.~\cite{Takahashi:2019ghj},
we computed the color correlation between a static quark and
antiquark ($q\bar q$) pair by means of an entanglement entropy (EE)
defined by the reduced density matrix $\rho$ in the color space
at zero temperature.
EEs defined by the reduced density matrix are known to quantify
correlations (i.e., entanglements) between
degrees of freedom in purely quantum systems,
and they have been used in many physical systems
~\cite{Itou:2015cyu,Aoki:2015bsa,Kanada-Enyo:2015ncq,Takayanagi:2012kg,Bennett:1995tk, Calabrese:2004eu, Vidal:2002rm, Amico:2007ag, Horodecki:2009zz, Bennett:1996gf, Wootters:1997id, Vidal:2002zz}.
Based on the EE as well as the reduced density matrix $\rho$,
color correlation was obtained as a function of the interquark
distance $R$~\cite{Takahashi:2019ghj}.
When totally color singlet quark and antiquark are located near each other,
the $q\bar q$ pair
forms a pure color singlet representation $|{\bm 1}\rangle$,
that yields the minimum EE (maximum color correlation).
As the $q\bar q$ separation $R$ is enlarged,
an octet color representation $|{\bm 8}\rangle$
is randomly mixed in, and as a result,
the quark color correlation is weakened,
indicating that even in the confined phase
a color screening between a $q\bar q$ pair occurs
in the large $R$ region.
If the color charge of the quark part 
and that of the gluon part are separately considered, 
such a color screening effect can be
understood as color leak from quarks to gluons.
The color charge flows from a $q\bar q$ pair
into in-between gluons (e.g. flux tubes)
as quark separation is increased~\cite{Tiktopoulos:1976sj, Greensite:2001nx}.
This color transfer from the quark part to the gluon part
is observed as a screening effect on quark color correlation.

An investigation based on a color density matrix $\rho$ and 
an EE $S$
enables a model-independent analysis of the color structure of quarks.
In this paper, we extend our previous work for zero-temperature systems
in Ref.~\cite{Takahashi:2019ghj}
to finite-temperature systems
and investigate the color correlation between a $q\bar q$ pair 
below and above $T_c$.
We define the reduced density matrix $\rho$ for a static $q\bar q$
pair in terms of color degree of freedom (DoF) at finite temperatures,
and compute an EE $S$ from the reduced density matrix $\rho$.
The reduced density matrix defined in the 
subspace of $q\bar q$ color configurations
is computed by integrating the gluons' degrees of freedom,
which can be achieved simply by averaging the density matrix components
over gauge configurations.
Through the analysis of $S$ and $\rho$~\cite{Takahashi:2019ghj},
we investigate the dependence of the color correlation
on the interquark distance $R$ and
the temperature $T$ below and above $T_c$.

In Sec.~\ref{Sec.Formalism}, we present the formalism necessary to compute
the reduced density matrix $\rho$ of the $q\bar q$ system
and its EE $S$. The details of the numerical calculations and the ansatz for $\rho$
are shown in Sec.~\ref{Sec.Formalism}.
The results are presented in Sec.~\ref{Sec.Results}, and
Sec.~\ref{Sec.Summary} is devoted to a summary and concluding remarks.

\section{Formalism}
\label{Sec.Formalism}

\subsection{Reduced 2-body density matrix and entanglement entropy}

In this paper, we consider the SU(3) Yang-Mills theory on the
lattice at finite temperature, and
follow the formalism shown in Ref.~\cite{Takahashi:2019ghj}.
The reduced two-body density operator $\hat\rho(R)$ in a $q\bar q$ system
with the interquark distance $R$ is defined as
\begin{equation}
\hat\rho(R) = {\rm Tr}_G\left(|q(0) \bar q(R)\rangle \langle q(0) \bar q(R)|\right),
\end{equation}
where $|q(0) \bar q(R)\rangle$ denotes a quantum state
in which an antiquark is located 
at the origin, while the other quark lies at $x=R$.
${\rm Tr}_G$ denotes the partial trace over the gluonic DoF.
The reduced density matrix components $\rho(R)_{ij,kl}$
with $i$, $k$ ($j$, $l$) being the quark's (antiquark's) color indices
are expressed as
\begin{equation}
\rho(R)_{ij,kl} = \langle q_i(0)\bar q_j(R)|\hat\rho(R)|q_k(0)\bar q_l(R) \rangle.
\end{equation}
Here, $\rho(R)$ is an $m\times m$ square matrix with dimension $m=N_c^2$,
and we directly compute $\rho(R)_{ij,kl}$ using lattice QCD techniques
shown in the next section.

The von Neumann EE $S^{\rm VN}(R)$ 
for a $q\bar q$ pair separated by $R$
is constructed from the reduced density matrix $\rho(R)$ as
\begin{eqnarray}
S^{\rm VN} (R)
&\equiv&
-{\rm Tr} \ \rho(R) \log\rho(R).
\end{eqnarray}
Another kind of entropy, the Renyi EE ~\cite{Renyi:1970}
$S^{{\rm Renyi}-\alpha}(R)$
of the order $\alpha$
is defined as
\begin{eqnarray}
S^{{\rm Renyi}-\alpha}(R)
&\equiv&
\frac{1}{1-\alpha}\log {\rm Tr}\left(\rho^\alpha \right).
\end{eqnarray}
When computing $S^{\rm VN}$,
one needs to diagonalize $\rho$ or approximate the logarithmic function,
and to avoid such numerically demanding processes,
throughout this paper,
we adopt Renyi entropy~\cite{Renyi:1970} at $\alpha =2$ for EE.
Note that in the limit when $\alpha\rightarrow 1$,
it goes to von Neumann entropy as $S^{{\rm Renyi}-\alpha}\to S^{\rm VN}$. 
Since the EEs, whose color indices are all contracted,
are invariant under unitary transformations,
they enable representation independent analysis
of the $q\bar q$ pair color correlation.

%%%%%%%%%%%%%%%%%%%

Here, we briefly explain the relationship between the $q\bar q$ color correlation
and the EE.
Letting $Q$ denote the quark and antiquark's color DoF
and $G$ denote the gluon's color DoF,
the $q\bar q$ pair's color correlation is defined in the subsystem $Q$.
A pure state in the entire $Q+G$ system,
which is created by a color singlet operator,
can be written as
\begin{eqnarray}
\sum_\alpha |\alpha\rangle_Q \otimes |\bar\alpha\rangle_G.
\end{eqnarray}
Here, $\alpha$ denotes all the possible color
representations of the $q\bar q$ pair,
and $\bar\alpha$ represents the color representation of the gluon field
which is determined so that $|\alpha\rangle_Q \otimes |\bar\alpha\rangle_G$
forms a color singlet state.
When the quark's and antiquark's colors 
are strongly correlated forming a color-singlet combination $|{\bm 1}\rangle_Q$
with
no color charge leak from $Q$ to $G$, 
the subsystems $Q$ and $G$ are decoupled in the color space, hence,
the whole state can be expressed as a direct product of $Q$ and $G$:
\begin{eqnarray}
%\sum_{\alpha={\bm 1}}^{\bm 1} |\alpha\rangle_Q \otimes |\bar\alpha\rangle_G
%=
|{\bm 1}\rangle_Q \otimes |{\bm 1}\rangle_G.
\end{eqnarray}
In this strongly correlated limit,
the EE $S^{\rm EE}$ goes to zero,
since the two subsystems $Q$ and $G$ 
demonstrate no entanglement with each other.
At the same time,
when the $q\bar q$ pair's color charge flows into in-between gluons,
and the $q\bar q$ color correlation decreases,
the entire state cannot be written in a separable form,
and $S$ becomes a positive finite value ($S>0$).

%%%%%%%%%%%%%%%%%%%

\subsection{Ansatz for reduced density matrix $\rho_{ij,kl}(R)$}
\label{ansatz}

In later sections,
we perform an analysis of the EE as well as the density matrix $\rho$
based on the ansatz proposed in Ref.~\cite{Takahashi:2019ghj}.
For the reader's convenience, here, we revisit the ansatz
that reproduces the lattice QCD data well.
We denote the density operator $\hat\rho_{{\bm s},{\bm s}}$
for the quark and antiquark forming a pure color singlet state
$|{\bm s}\rangle = \sum_i^{N_c} |\bar q_i q_i\rangle$
in the Coulomb gauge as
\begin{equation}
\hat\rho_{{\bm s},{\bm s}} = |{\bm s}\rangle \langle {\bm s}|.
\end{equation}
The density operator $\hat\rho_{{\bm a}_i,{\bm a}_i}\ \ (i=1\sim N_c^2-1)$
for $q\bar q$ in an adjoint state 
$|{\bm a}_i\rangle\ \ (i=1\sim N_c^2-1)$
in color SU($N_c$) QCD
is expressed as
\begin{equation}
\hat\rho_{{\bm a}_i,{\bm a}_i} = |{\bm a}_i\rangle \langle {\bm a}_i|
\ \ (i =1\sim N_c^2-1).
\end{equation}
We assume that 
random color configurations of the $q\bar q$ pair,
without any color correlation between quark and antiquark,
are mixed with the maximally correlated color singlet component
as the interquark distance $R$ is increased due to QCD interaction.
In our previous paper, we demonstrated that this assumption is satisfied at $T=0$.
In this ansatz, the density operator $\hat \rho$ can be written as
\begin{eqnarray}
 \hat\rho_{\rm ansatz}(R)
  &=&
  F(R)\hat\rho^{\rm 0}+(1-F(R))\hat\rho^{\rm rand}.
\end{eqnarray}
Here, $F(R)$ represents the fraction of the original (maximally correlated)
singlet state
and $(1-F(R))$ is that of the mixed (random) components.
The maximally correlated state in which
the $q\bar q$ pair forms $|{\bm 1}\rangle$
gives $F(R)=1$,
and at the random limit, when the quarks' colors are completely screened,
it is represented by $F(R)=0$.

The explicit forms of $\hat \rho^{\rm 0}$ and $\hat \rho^{\rm rand}$ 
are given as
\begin{eqnarray}
&&
 \hat \rho^{\rm 0}
=
 \hat\rho_{{\bm s},{\bm s}}
\\
&&
 \hat \rho^{\rm rand}
=
 \frac{1}{N_c^2}\hat\rho_{{\bm s},{\bm s}}
 +
 \frac{1}{N_c^2}\hat\rho_{{\bm a}_1,{\bm a}_1}
 +
 \frac{1}{N_c^2}\hat\rho_{{\bm a}_2,{\bm a}_2}
 +...
\end{eqnarray}
When $N_c=3$, this ansatz implies
\begin{eqnarray}
 \begin{cases}
 \rho(R)_{{\bf 8}_1,{\bf 8}_1}
 =
 \rho(R)_{{\bf 8}_2,{\bf 8}_2}
 =...=
 \rho(R)_{{\bf 8}_8,{\bf 8}_8}
\equiv
 \rho(R)_{{\bf 8},{\bf 8}}
\\  
 \rho(R)_{\alpha,\beta}=0\ \ ({\rm for}\ \alpha\neq\beta)
 \end{cases}
\label{Eq.conditions}
\end{eqnarray}
The first relation should be satisfied due to the color SU(3) symmetry.
The normalization condition ${\rm Tr}\rho = 1$
is trivially satisfied in this ansatz as
\begin{eqnarray}
 \rho(R)_{{\bf 1},{\bf 1}}
+
(N_c^2-1)
 \rho(R)_{{\bf 8},{\bf 8}}
=1.
\label{Eq:normalization}
\end{eqnarray}
In Ref.~\cite{Takahashi:2019ghj},
the density matrix $\rho(R)_{\alpha,\beta}$ obtained by the lattice calculation
was found to satisfy Eq.(\ref{Eq.conditions})
with good accuracy for all $R$.
Thus, we can perform analyses based on $\rho_{\rm ansatz}(R)_{\alpha,\beta}$
instead of $\rho(R)_{\alpha,\beta}$,
and, hereafter, we omit the subscript ``ansatz'' in $\rho_{\rm ansatz}(R)_{\alpha,\beta}$.

Taking into account the normalization condition,
the independent quantity 
at a given $R$ in this ansatz
is only $\rho(R)_{{\bf 8}, {\bf 8}}$
in Eq.(\ref{Eq.conditions}),
which we obtain as the averaged value of the lattice density matrix
elements,
$\rho(R)_{{\bf 8}, {\bf 8}}=\frac{1}{N_c^2-1}\sum_i \rho(R)_{{\bf 8}_i, {\bf 8}_i}$.
Then, we can compute the fraction $F(R)$ of the remaining correlated $q\bar q$
component as,
\begin{equation}
 F(R) 
 =\rho(R)_{{\bf 1},{\bf 1}}-\rho(R)_{{\bf 8},{\bf 8}}
= 1-N_c^2 \rho(R)_{{\bf 8},{\bf 8}}.
\label{Eq.F(R)}
\end{equation}

\subsection{Lattice QCD setup}

We define a Polyakov line at the spatial point ${\bm x}$ as
\begin{eqnarray}
 P_{ij}({\bm x})\equiv
\left[\prod_{\tau=0}^{N_\tau -1} U_4({\bm x},\tau)\right]_{ij},
\end{eqnarray}
where $i$, $j$ denotes the color index, and $\tau$ is the temporal position.
$U_4({\bm x},\tau)$ is a Euclidean 
temporal link variable at $({\bm x}, \tau)$.

With a normalization factor $N$ determined such that
\begin{eqnarray}
{\rm Tr}\ \langle \frac{1}{N}P^\dagger_{ij}(0)P_{kl}(R) \rangle =\sum_{ij} \langle \frac{1}{N}P^\dagger_{ij}(0)P_{ij}(R) \rangle = 1,
\end{eqnarray}
we obtain $\rho(R)_{ij,kl}$ whose trace is unity (${\rm Tr}\ \rho(R)=1$)
as 
\begin{eqnarray}
\rho(R)_{ij,kl} = \langle \frac{1}{N}P^\dagger_{ij}(0)P_{ij}(R) \rangle.
\end{eqnarray}

Once we obtain $\rho(R)$, Renyi entropy of the order $\alpha$
as a function of $R$ can be obtained as
\begin{equation}
S^{{\rm Renyi}-\alpha}(R) = \frac{1}{1- \alpha}\log {\rm Tr}\left(\rho(R)^\alpha \right).
\end{equation}

\begin{figure}[h]
\includegraphics[width=08cm]{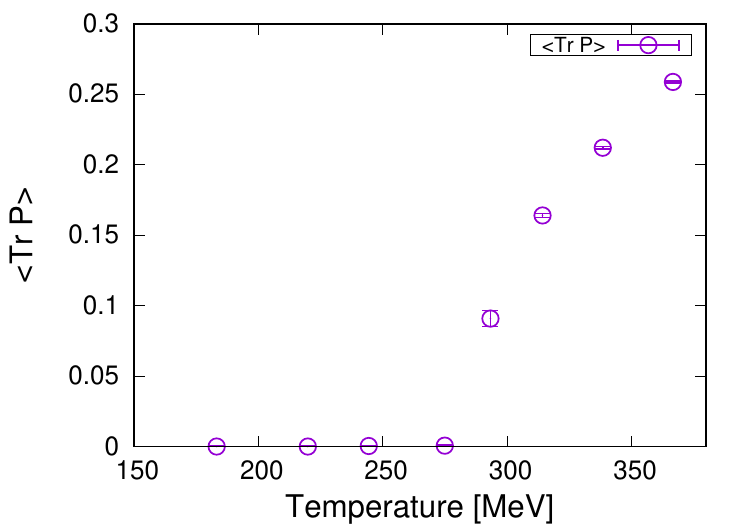}
 \caption{\label{PasT}
The expectation value of a Polyakov loop $\langle {\rm Tr}P\rangle$
plotted as a function of the temperature $T.$}
\end{figure}

In this study, we adopt the standard Wilson gauge action
that exhibits a temporal anisotropy,
and perform quenched calculations for reduced density matrices
of static $q\bar q$ systems at finite temperatures.
The gauge configurations are generated
on the spatial volume $L^3 = 24^3$, 
with the gauge coupling $\beta = 5.75$ and the renormalized temporal anisotropy
$\xi = 4.0$~\cite{Klassen:1998ua}.
This leads to a spacial cut off $a_\sigma^{-1}$ of 1.1 GeV and a temporal cut
off $a_\tau^{-1}$ of 4.4 GeV~\cite{Matsufuru:2001cp}.
The temporal extents $N_\tau$ employed in this work
are $N_\tau = 12, 13, 14, 15, 16, 18, 20$ and $24$,
which correspond to the temperatures of 183--367 MeV.
All the gauge configurations are gauge-fixed with the Coulomb gauge condition.
The deconfinement transition was investigated using an anisotropic
lattice and found to occur
at $T=T_c=285$ MeV~\cite{Meng:2009zzb}.
Since a precision study of the transition temperature $T_c$ is not
within the scope of the present work,
we conclude this section simply by
showing the expectation value of a Polyakov loop 
$\langle {\rm Tr}P\rangle$,
an order parameter in quenched QCD,
obtained in our setup in Fig.~\ref{PasT}.

\section{Lattice QCD results}
\label{Sec.Results}

\subsection{Entanglement entropy}

We adopt Renyi entropy of $\alpha=2$ for the evaluation of the color correlation between $q\bar q$ pairs.
The $S^{\rm Renyi-2}(R)$ is obtained by taking
the logarithm of the trace of the squared reduced density matrix $\rho(R)$
as
\begin{equation}
S^{{\rm Renyi}-2} =-\log {\rm Tr}(\rho(R)^2).
\end{equation}
Taking into account the normalization condition
${\rm Tr}(\rho(R))=1$, 
the maximum of $S^{{\rm Renyi}-2}$ is reached when
all the $N_c^2$ diagonal elements are equal to $1/N_c^2$
in the diagonal representation of $\rho(R)$.
Based on the representation invariance of $S$,
the maximum value of $S$ is
\begin{equation}
{\rm max}\left[S^{\rm Renyi-2}(R)\right]
=2\log N_c.
\end{equation}

\begin{figure}[h]
\begin{center}
\includegraphics[width=08cm]{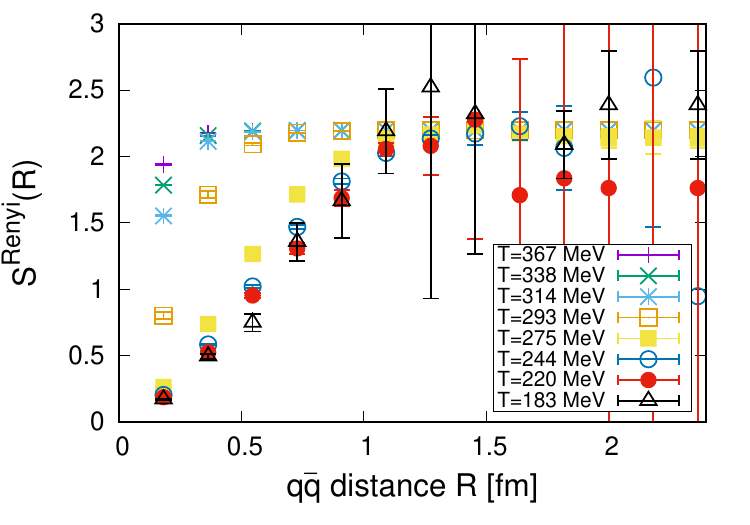}
\includegraphics[width=08cm]{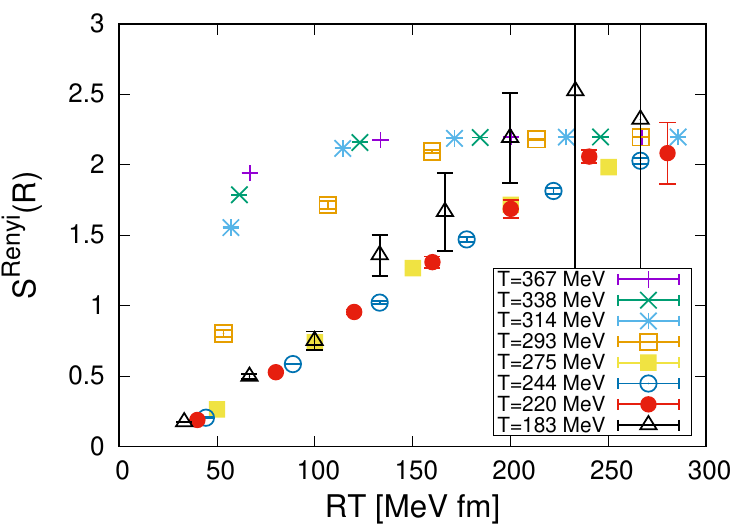}
\end{center}
 \caption{\label{SRenyi2}
 The upper panel shows $S^{\rm Renyi-2}_{\rm lattice}(R)$ at each $T$ obtained
 from the reduced density matrix $\rho(R)$
 plotted as a function of the interquark distance $R$,
 and the lower panel indicates $S^{\rm Renyi-2}_{\rm lattice}(R)$ plotted against $RT$.
 }
\end{figure}

In the upper panel in Fig.~\ref{SRenyi2}, $S^{\rm Renyi-2}(R)$ values calculated
from the $\rho(R)$ obtained on the lattice are plotted.
At small $R$ values, $S^{\rm Renyi-2}(R)$ exhibits smaller values
indicating that the $q\bar q$ pair forms a pure color singlet state
and is strongly correlated at $R\rightarrow 0$.
$S^{\rm Renyi-2}(R)$ rises with increasing $R$ and
reaches a maximum value ($2\log N_c$)
at a large $R$ value, indicating that
the $q\bar q$ color configuration approaches a random configuration
without color correlation in the large $R$ range.
This tendency is observed at all temperatures.

Below $T_c=285$ MeV, $S^{\rm Renyi-2}(R)$ values are more or less 
similar to each other,
indicating that the thermal effect on the color correlation
in the confined phase is not apparent in investigated temperature range.
At $T=T_c$, the $q\bar q$ color correlation is quickly quenched 
due to the deconfinement transition,
which can be observed as a sudden increase in the $S^{\rm Renyi-2}(R)$
across $T_c$.
Above $T_c$, $S^{\rm Renyi-2}(R)$ for $R\ge 0.6$ fm
take almost the maximum value 
and reveal a vanishing $q\bar q$ color correlation.
In contrast,
$S^{\rm Renyi-2}(R)$ at $R< 0.6$ fm have
a $T$ dependence,
and it shows that the color correlation 
still exists between the $q\bar q$ pair at short distances
even in the deconfined phase.

To take a brief look at the thermal effect on the $q\bar q$ color correlation,
we plot $S^{\rm Renyi-2}$ as a function of $RT$ in the lower panel in
Fig.~\ref{SRenyi2}.
It is found that the EEs above $T_c$ ($=285$ MeV) lie on a common curve,
except for those for $T=293$ MeV just above $T_c$,
which is considered to be transient region.
This indicates that the color screening effect above $T_c$
caused by thermal fluctuations depends on $R$ normalized by $1/T$,
which would be related to the Debye screening length in the deconfined phase.
We discuss the color screening mass in a quantitative way in later sections.
It is interesting that the EEs obtained below but near $T_c$
also lie on a common curve.

\begin{figure}[h]
\begin{center}
\includegraphics[width=08cm]{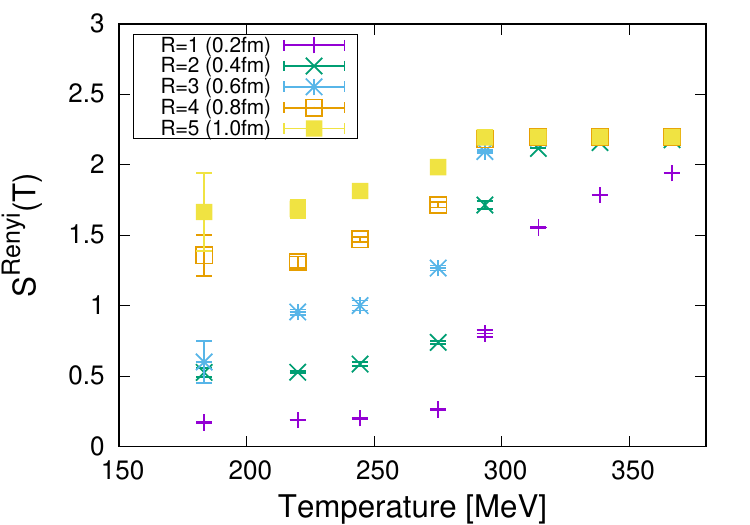}
\end{center}
 \caption{\label{SRenyi2asT}
$S^{\rm Renyi-2}_{\rm lattice}(R)$ at each $R$
 plotted as a function of the temperature $T$.}
\end{figure}

To investigate the $T$ dependence of EE in more detail
at each distance $R$, we plot 
the EEs as a function of $T$ (Fig.~\ref{SRenyi2asT}).
Below $T_c$, 
the $S^{\rm Renyi-2}(T)$ values for $R<0.6$ fm are less than 30\% of the maximum value,
revealing the strong correlation between a $q\bar q$ pair.
For $R\geq 0.6$ fm, $S^{\rm Renyi-2}(T)$ reach 
greater than 50\% of the maximum value, indicating a reduced color correlation
due to color leak to the in-between flux-tube.
At $T=T_c$, the EEs for all $R$ increase suddenly,
and the $q\bar q$ color correlation is
mostly lost after the deconfinement phase transition.
Taking into account that the quarks' color correlation is closely
related to the surrounding gluon configuration,
the increase in EE at $T=T_c$ reflects a drastic modification
of the gluon configuration
at the phase transition.
Above $T_c$, 
the EEs for $R< 0.6$ fm do not reach the maximum value at $T$ values close to $T_c$,
continuing to grow as $T$ increases. 
This indicates that, even in the deconfined phase,
a thermal effect on the $q\bar q$ color correlation still exists.
This behavior can be understood based on the thermal effect 
on color screening by the gluon medium,
which will be discussed in the 
the context of the "color screening mass" in a later section.

While EEs show a sudden increase at the phase transition point for any $R$ ,
the EE at the shortest distance ($R=0.2$ fm)
shows an especially clear change and behaves like an order parameter,
the expectation value of a Polyakov loop (Fig.~\ref{PasT}).
In the limit $R\rightarrow 0$, the $q\bar q$ pair is dominated by
a strongly correlated color singlet state $|{\bm 1}\rangle$
and no color leak to the system 
occurs due to the color confinement,
which leads to the small EE, amounting to just 10\% of the maximum value
at $T<T_c$.
After the deconfinement phase transition,
$q\bar q$ pair is also able to form a color non-singlet (octet) state $|{\bm 8}\rangle$
even in the case of $R\rightarrow 0$.
In this case, the color charge largely spread into the surrounding gluon field
and the color correlation is lost.
This sudden mixture of the color octet $q\bar q$ states
leads to significant enhancement of EE above $T_c$.

It should be noted that below $T_c$,
the $T$ dependence of $S^{\rm Renyi-2}_{\rm lattice}(R)$
is small for all $R$ values,
but the $S^{\rm Renyi-2}_{\rm lattice}(R)$ is not entirely constant,
as can be seen in Figs.~\ref{SRenyi2} and \ref{SRenyi2asT}.
This may indicate a thermal influence on the color correlation 
in the confined phase.
For a clarification of the thermal effects in the confined phase,
a further detailed analysis with high statistics is needed.

\subsection{Analysis using an ansatz}
\label{Sec.ansatz}

To investigate the $q\bar q$ color correlations 
in more detail,
we compute $F(R)$ defined in Sec.\ref{ansatz} 
by adopting the ansatz Eq.(\ref{Eq.F(R)}).
The color screening effect by the in-between gluon field
is encoded in $F(R)$, which is the only parameter in this ansatz
that indicates the fraction of the color-singlet component
of the $q\bar q$ pair.
$F(R)=1$ for a pure color singlet state
(the maximally correlated $q\bar q$ pair),
and $F(R)=0$ for the random color state of the $q\bar q$ pair
(the uncorrelated $q\bar q$ pair). 
\begin{figure}[h]
\begin{center}
\includegraphics[width=08cm]{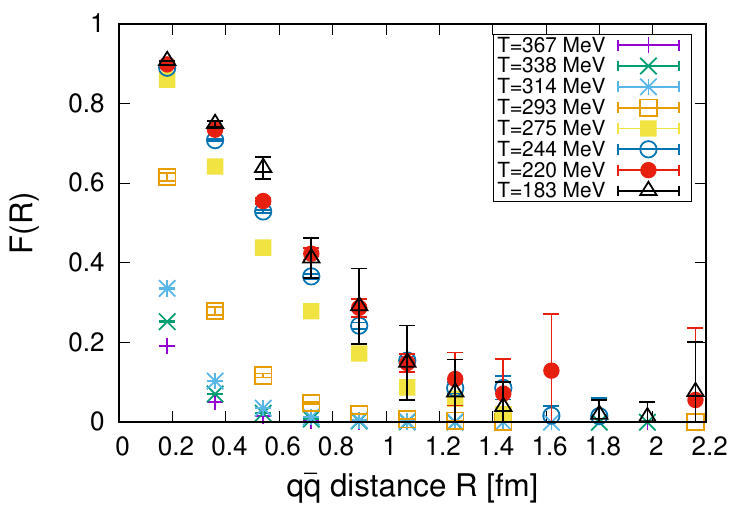}
\includegraphics[width=08cm]{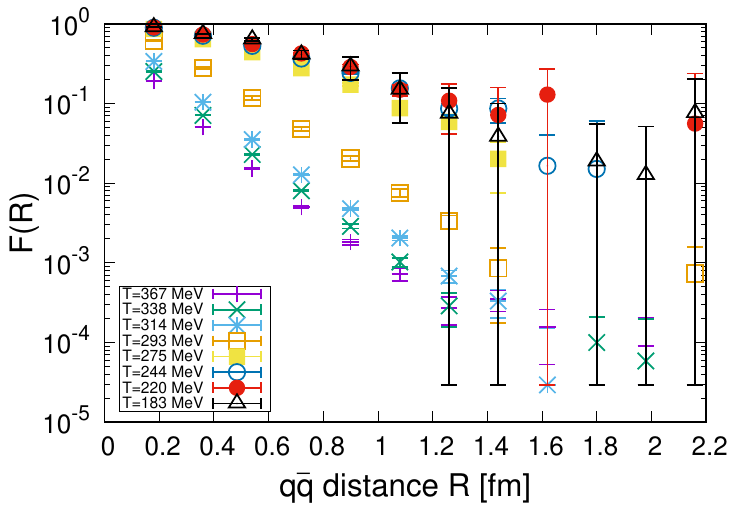}
\end{center}
 \caption{
 \label{F(R)}
 The upper panel shows $F(R)$ at each temperature 
 plotted as a function of the interquark
 distance $R$. $F(R)$ monotonously decreases,
 approaching zero. The lower panel shows a logarithmic plot of $F(R)$.
 }
\end{figure}

In Fig.~\ref{F(R)},
$F(R)$ at each $T$ is plotted as a function of the interquark
distance $R$. 
The lower panel in Fig.~\ref{F(R)} shows the logarithmic plot of $F(R)$.
Below $T_c$,
$F(R)$ appears to change its slope around $R=0.6$ fm at each $T$,
as can be seen in the lower panel.
$F(R)$ seems to linearly decrease in the short-distance region 
($R<0.6$ fm), while decaying exponentially at larger $R$ values ($R\geq 0.6$ fm).
This change in $R$ dependence might indicate
flux-tube formation between $q\bar q$ pairs at $R\geq 0.6$ fm.
In other words,
a flux tube is generated between $q\bar q$ pairs with a large separation $R$,
and the in-between gluons forming the flux tube screen the
quarks' color, leading to exponential decay of the correlation.
At the same time, in the small $R$ region,
the quarks' color is not screened by surrounding gluons,
which instead produce Coulomb-type interactions.
To investigate this in detail, here, we analyze 
the damping factor in these two distinct regions of $R$,
defining a "color screening mass" in the next section. 
Above $T_c$, 
$F(R)$ appears to exponentially decay over all $R$ regions.
This damping behavior can be understood 
as a medium effect in the deconfined phase, 
where a $q\bar q$ pair feels a screened Yukawa-type potential 
over the entire $R$ range;
hence, $F(R)$ provides exponential damping with a constant damping factor. 

\subsection{Color screening mass}

The exponential damping of the $q\bar q$ correlation $F(R)$ indicates
the color screening effects of in-between gluons.
We fit $F(R)$ with an exponential function as
\begin{equation}
F(R)  = A \exp(-B R),
\end{equation}
and extract the ``color screening mass'' $B$.
In particular,
since the behavior of $F(R)$ appears to qualitatively change at $R\sim 0.6$ fm,
we adopt two distinct fit ranges
($R < 0.6$ fm and $R \geq 0.6$ fm).
\begin{figure}[h]
\includegraphics[width=08cm]{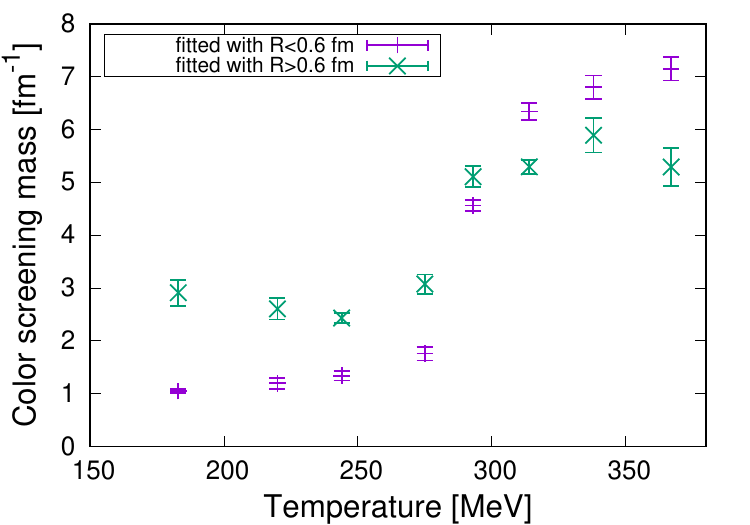}
 \caption{\label{scmass}
The color screening masses $B$ 
fitted in the range of $0\leq R \leq 0.6$ and $R \geq 0.6$ fm
plotted as a function of $T$.
The fit was performed with the functional form $F(R)=A\exp(-BR)$.}
\end{figure}
In Fig.~\ref{scmass},
the color screening masses $B_<(T)$ and $B_>(T)$ extracted 
in the two ranges
$R < 0.6$ and $R\geq 0.6$ fm
are plotted as a function of the temperature $T$.

Below $T_c$, 
$B_>(T)$ and $B_<(T)$ show a small $T$-dependence,
implying that the thermal effect
on the color screening factor is not high
in the $T$ range investigated in the present analysis.
$B_>(T)$ is approximately two times larger than $B_<(T)$,
with this discrepancy
indicating that the surrounding gluon configuration in the short ($R<0.6$ fm) and long ($R\geq 0.6$ fm) range regions,
which are responsible for the screening mass,
differ from each other in the confinement phase.
While the gluon field stemming
from a $q\bar q$ pair with a small separation $R$
yields a Coulomb type potential,
it produces a linear confinement potential forming an in-between flux tube
in the large $q\bar q$ separation region.
A color leak to the flux tube leads to
a mixture of random color components and yields a large screening mass.
At $T=T_c$, both $B_>(T)$ and $B_<(T)$ 
show a sudden increase to much higher values,
and this sudden change in the screening mass reflects the
deconfinement phase transition, in which
the color flux configuration around the $q\bar q$ pair exhibits a drastic change.
Above $T_c$, $B_>(T)$ and $B_<(T)$ take more or less the same values,
and the ratio of $B_>(T)$ to $B_<(T)$ is close to one.
This is naturally expected, since the gluon field at $T>T_c$
produces a Yukawa-type potential for a $q\bar q$ pair over the whole $R$.
$B_<(T)$ seems to gradually increase above $T_c$,
which could be related to the increase of the screening mass
in the Yukawa-type potential in the deconfined phase.

The jump at $T=T_c$ can be found in both $B_>(T)$ and $B_<(T)$
also in the screening mass.
$B_<(T)$ is much smaller than $B_>(T)$ below $T_c$,
and an increase in $B_<(T)$ can be clearly observed.
This order-parameter-like behavior of the color screening mass $B_<(T)$
in the short range region can again be understood
as the sudden modification of the color flux configuration around the quarks.
When the interquark distance $R$ is sufficiently small at a lower temperature,
the color flux does not flow into a system around $q\bar q$
due to the color confinement,
according to which only a color singlet $q\bar q$ state is physical.
After the deconfinement transition occurs,
a $q\bar q$ pair is able to even form a color-octet state,
with the color largely leaking from the quarks throughout the system.
This color leak leads to the sudden change
of the color flux configuration, 
significantly increasing the color screening mass $B_<(T)$ across $T_c$.

Finally, here, we make a comparison between the screening masses
obtained in our analysis and those estimated in
previous works~\cite{Burnier:2016mxc,Kaczmarek:2005ui,Laine:2019uua}.
In Ref.~\cite{Burnier:2016mxc}, the Debye mass $m_D$
in an in-medium heavy quark
potential was investigated and researchers found that $m_D/T\sim 1.2$ at $T/T_c=1.5$.
In Ref.~\cite{Kaczmarek:2005ui},
$m_D/T\sim 2.5$ was obtained
from an analysis of the color singlet $q\bar q$ free energy $F_1$.
The Debye mass $m_E$ estimated at 2-loop order shows that $m_E/T\sim 2.5$
in the temperature range we investigated here.
The ``color screening mass'' $B$ obtained in the current analysis
amounts to $B/T\sim 3.5$ at $T/T_c=1.5$,
which is significantly larger than the values found in previous studies.
While previous estimates of the Debye
mass~\cite{Burnier:2016mxc,Kaczmarek:2005ui} were based on the heavy
quark potential,
our screening mass was extracted from the color correlation function $F(R)$.
The relationship between the heavy quark potential and the color
correlation function leaves room for discussion. 
Another possible origin of the deviation may lie in the channel
investigated. 
While the target channel in
Refs.~\cite{Burnier:2016mxc,Kaczmarek:2005ui}
is a color singlet channel, our present calculation contains
``contaminations'' from the color octet states above $T_c$,
which could have caused the difference.

\section{Summary and concluding remarks}
\label{Sec.Summary}

In this study, by means of an EE
defined by reduced density matrices $\rho$ in a color space,
we have investigated the color correlation
of static $q\bar q$ systems at finite temperatures.
We have adopted the standard Wilson gauge action on an anisotropic lattice
and performed quenched calculations for density matrices.
The gauge coupling is $\beta = 5.75$ and
the spatial volume is $L^3 = 24^3$.
To directly evaluate the components of $\rho_{ij,kl}$,
we have imposed the Coulomb-gauge on the gluon field.
The temperatures we investigated range from 183 to 367 MeV.

Further, we have evaluated the $q\bar q$ correlation
as a function of the $q\bar q$ separation $R$
based on the EE constructed
from the reduced density matrix $\rho$ at finite temperatures.
The Renyi entropy $S^{{\rm Renyi}-\alpha}$ of the order $\alpha =2$ 
is adopted for the evaluation of EE.
Below $T_c=285$ MeV,
we have found that the EE at each $q\bar q$ separation $R$
is barely affected by temperature,
showing similar behavior to that at $T=0$.
When a $q\bar q$ pair is located nearby, it forms a strongly correlated
(color singlet) state $|{\bm 1}\rangle$,
and random color components (a random mixture of $|{\bm 1}\rangle$
and $|{\bm 8}\rangle$)
appear, depending on $R$ in the large $R$ region,
leading to color screening between quarks.
As the temperature increases, such $q\bar q$ correlation is quickly lost
during the phase transition
at $T=T_c$ leading to an acute increase of the EE,
which reflects the drastic modification
of the gluon configuration at the phase transition.
While the EEs for $R\ge 0.6$ fm reach almost the maximum value
above $T_c$,
the EEs for $R< 0.6$ fm continue growing as $T$ increases,
indicating that the $q\bar q$ correlation
has a temperature dependence at $T>T_c$.

To understand the $R$ dependence of the $q\bar q$ color correlation in detail,
we have extracted the color correlation function $F(R)$
based on the ansatz for the reduced density matrix $\rho$,
in which $\rho$ is written by a sum of
the color-singlet (correlated) state $|{\bm 1}\rangle\langle{\bm 1}|$
and random uncorrelated
elements $|{\bm 1}\rangle\langle{\bm 1}|$,
$|{\bm 8_i}\rangle\langle{\bm 8_i}|$ ($i=1,..,N_c^2-1$).
In the analysis, we have extracted the ``color screening mass'' $B$
from the color correlation function $F(R)$,
which represents the strength of the color screening.
In particular, we have evaluated the short-range screening mass $B_<$
and the long-range mass $B_>$ defined in the ranges $R<0.6$ and $R\geq 0.6$ fm,
respectively.
Below the phase transition temperature $T_c$, $B_<$ and $B_>$ differ in value,
indicating the qualitative difference in gluon field profile
between these two distinct ranges:
One demonstrates a Coulomb-type gluon configuration
and the other demonstrates a one-dimensional flux-tube profile.
At the phase transition point $T=T_c$, 
the screening mass significantly and suddenly increases.
Taking into account that the color screening mass encodes
the surrounding gluon configuration around the $q\bar q$ pair,
the enhancement of $B$ at $T_c$ should be related 
to the color flux delocalization caused by the deconfinement transition.
Above $T_c$, $B_<$ and $B_>$ take similar values,
consistent with the gluon profile,
which gives a Yukawa-type screened potential over the entire $R$ range.

The EE and the screening mass values have been found
to significantly change
at the phase transition point, indicating that
the EE properly detects the significant modification 
of the gluon configuration around the $q\bar q$ pair
across the color deconfinement transition.
In particular, the EE and the screening mass for small $R$ values
behave like order parameters of the color confinement.
Though these are not order parameters
related to QCD symmetry,
they demonstrate a clear physical meaning and can be good probes
for the phase transition in finite $T$/$\mu$ systems.

In conclusion, the $q\bar q$ color correlations
have been found to be well quantified by EEs
at finite temperatures below and above $T_c$.
These findings encourage us 
to adopt EE for the future study of the internal color structures of hadrons
including multiquark systems~\cite{Okiharu:2004ve}
or the color structures of systems at finite $T$/$\mu$.

\begin{acknowledgments}
This work was partly supported
by Grants-in-Aid of the Japan Society for the Promotion of Science 
(Grant Nos. JP18K03617 and JP18H05407).
\end{acknowledgments}

%\newpage

\end{document}